# Mixing Strategies in Cryptocurrencies and An Alternative Implementation

Xinyuan Zhang

**Abstract.** Since the initial launch of Bitcoin[3] by Satoshi Nakamoto in 2009, decentralized digital currencies have long been of major interest in both the academia and the industry. Till today, there are more than 3000 different cryptocurrencies over the internet[4]. Each one relies on mathematical soundness and cryptographic wit to provide unique properties in addition to securing basic correctness. A common misbelief for cryptocurrencies is that they provide complete anonymity by replacing people's real-life identity with a randomly generated wallet address in payments. However, this "pseudonymity" is easily breakable under the public ledger. Many attacks demonstrate ways to deanonymize people through observing the transaction patterns or network interactions. Thus, cryptocurrency fungibility has become a popular topic in the research community. This report reviews a partial list of existing schemes and describes an alternative implementation, Eth-Tumbler. Eth-Tumbler utilizes layered encryption and multiple signatures and thus efficiently hides a user under k-anonymity

## I. Introduction

Fungibility is an important property for currencies. For instance, banknotes are interchangeable and indistinguishable. Every piece of paper money carries its value without any additional information about its former or current owners. In centralized e-cash protocols, banks act as a trusted intermediary that guarantees the untraceability of each transaction. Blockchain-based cryptocurrencies, such as Bitcoin, try to preserve user privacy by hiding the user's identity with a wallet address, which is the hash of a public key. Although one cannot directly deduce the relationship between an address and any individual, it is possible to make good guesses by examining the flow of money on the public ledger. Therefore, Bitcoin provides only

"pseudonymity" for its users. There are many attempts to enhance privacy guarantees through introducing more cryptographic primitives to the current system. These modifications are building blocks for more complicated protocols that are developed later:

**Confidential Transaction**

The confidential transaction[14] ensures the correctness of transactions while hiding the actual amount. This cryptographic protocol encrypts the value of a transaction and leverages the technique of additively homomorphic commitments to make validating the total amount still possible. Confidential transaction does not require additional basic cryptographic assumptions to the current Bitcoin system and has a manageable level of overhead.

**Hierarchical Deterministic Wallet**

The hierarchical deterministic wallet[21] derives a deterministic tree of public keys from a single seed as a starting point. Users can generate and recover the seed with a memorizable mnemonic. The seed is then used to derive a master key pair and the master keys can be used to compute the tree of child keys. This system allows users to not only effectively create and own multiple addresses, but also easily back up and restore a wallet. It also enables a user to calculate other users' addresses with a public key on a branch and without knowing any corresponding private keys. Therefore, the users can confuse others by changing the receiving address for each transaction and thus protect one's financial privacy by stopping some linking attacks. The BIP32 standard implements HD Wallets and is widely adopted by different Bitcoin wallets.

**Stealth Address**

Stealth address[2] requires payers to create a random address based on a public address of the payee and sends coins to that one-time destination directly. Stealth address aims to make it impossible to link transactions to the payee's published long-term address and the one-time generated addresses. Only the stealth address owners can use their private keys to retrieve incoming transactions. It is thus a method that provides additional security and privacy to payees without any communication between payers and recipients.

Even though a user can encrypt transaction amounts, own multiple addresses, and hide among newly generated addresses, an online address can still be potentially mapped to a real entity through analyzing network or observing wallet interactions. There are even companies, such as Elliptic[9] and Chainalysis[5], that audit transactions on the blockchain and try to detect illegal events. Therefore, a technique called mixing emerges and further minimizes the information revealed on the publicly accessible blockchain.

## II. Existing Works

Many existing works utilize the idea of mixing to examine and address this anonymity problem. Most of them fall in three main categories: peer-to-peer on-chain mixing, mixing with intermediary, and alternative coins. Peer-to-peer on-chain mixing ensures k-anonymity for a single round of mixing in a UTXO(unspent transaction output) model. It combines the transactions of a group to fit in one block and thus removes the link between coin senders and output receivers. Mixing with intermediary relies on a third party to facilitate the protocol. This third party can be partially trusted or even trustless and this method often has better scalability. Alternative coins are new kinds of cryptocurrency and usually provide better anonymity guarantees. Below I summarize and analyze each category with more details and some examples:

**Peer-to-peer On-chain Mixing**

CoinJoin[12] is one of the first mixing ideas proposed in 2013 by Dr. Gregory Maxwell and it has become the foundation of many later protocols. It is a general idea that people can form a group and perform a joint transaction to break the link between senders and receivers on the blockchain. It has many possible implementations. The simplest version would reveal everyone's destination among senders, while more complicated versions utilize zero-knowledge proofs and provide better anonymity but are significantly more expensive.

There are many implementations and variations of CoinJoin. One of the implementations in use is called JoinMarket[18]. JoinMarket introduces a maker-taker concept for people to either form new groups and wait for enough people or join an existing group with the ideal parameters for payments, such as time to complete transaction or group denomination for transaction. Another implementation that is based on CoinJoin is CoinShuffle[25]. CoinShuffle combines a technique called layered encryption with mixing to provide cheap but better privacy among senders. The layered encryption works as the following. Users broadcast their individual public keys and each encrypts their destinations under all public keys. They can then form a chain of communication, decrypt one layer of all destinations one at a time, and shuffle the result before posting to the public channel until the plaintext destinations are revealed to everyone in the channel. Coinshuffle guarantees payer anonymity because no one gets to see the link between a sender and a destination address. Its extended version ValueShuffle[24] adds Confidential Transaction for value privacy and Stealth Address for payee anonymity. ValueShuffle no longer requires premixing or payer-payee interactions.

Peer-to-peer on-chain mixing protocols are compatible with the Bitcoin system and many other cryptocurrencies that use UTXO models. So they are cheap to implement. They are also usually effective and user-friendly. However, there are many problems beside this convenience. Most of them are vulnerable to DoS attacks and Sybil attacks because they rely on a group of other people to successfully complete one transaction. DoS attack is when a user is forced to be excluded from the protocol. Sybil attack happens when a large fraction of users in a group is compromised by a single entity and the real anonymity set gets significantly reduced. XIM[11] is proposed to disincentivize Denial-of-Service and Sybil attacks by adding a fee for mixing. It also utilizes the blockchain to broadcast transaction information in order to find a suitable partner to mix together. But XIM is only compatible with a group of two. What's more, all the protocols have very limited scales because of the complexity caused by multiple rounds of communication required to form a group and share data. To avoid communication and time complexity caused by waiting for a right group, many schemes start to introduce a third-party for handling the transactions.

**Mixing with Intermediary**

CoinSwap[13] is an early mixing idea with an intermediary, again proposed by Dr. Gregory Maxwell. It realizes coin exchanges with no risk of theft. The main building block is a trusted intermediary, which simulates two correlated 2-of-2 escrows. At the high level, this escrow first waits for the sender's money for a specific time window. Once the sender fulfills the payment, it locks the fund for another duration and the coins can only be released to the recipient with both parties' signatures. If no correct action is taken within the time window, the payer can reclaim the coins. The intermediary cannot steal because of a technique called hashlock transaction. What's more, the intermediary behaves correctly because the payer only releases the fund if the recipient gets paid. One advantage of CoinSwap is that it makes peer-to-peer cross-chain trade possible, while CoinJoin is assumed to happen on a single chain. But one problem with CoinSwap is its efficiency. The intermediary does not eliminate off-chain communication costs and the protocol requires at least 4 confirmations on the blockchain. Another main problem is that the link between sender and receiver is revealed to the escrow.

Another scheme that has the exact same problem of anonymity to only external observers is MixCoin[17]. MixCoin adds a warranty for payments to incentivize accountable intermediary. If the mixer steals funds, the payer can prove that the mixer is malicious and hence destroy the mixer's reputation. So a mixer can steal, but it cannot get away with the cheating. It is slightly cheaper than CoinSwap since it requires 2 block confirmations instead of

4. But it adds a mixing fee to also incentivize honesty for participants. In addition, the mixer is a single point of failure for DoS attacks.

Both of the above protocols expose participants' identities to the central mixer. A later attempt called TumbleBit[8] thoroughly resolves this problem. The intermediary no longer needs to be trusted. TumbleBit relies on a centralized tumbler that uses RSA puzzles to prevent stealing and to provide unlinkability. There are three phases. In phase one, the escrow phase, both payer and payee set up payment channels with the tumbler respectively and the payee obtains an RSA puzzle as a promise for payment from the tumbler. In phase two, the payment phase, the payer solves the puzzle by interacting with the tumbler and hands the solution to the payee. In phase three, the cash-out phase, the payee submits a transaction created from the puzzle solution and obtains the escrowed payment from the tumbler. TumbleBit supports a significantly larger scale of user size with off-chain payment. It also requires much less coordination among users. Payers and payees only need to interact with each other and the tumbler. A unique problem with TumbleBit is that there have to be enough coins in the tumbler to begin with in order to set up the escrow. What's more, the protocol will fail and the payer-payee link will be revealed when the tumbler colludes with a TumbleBit user.

**Alternative Coins**

Protocols in both peer-to-peer on-chain mixing and mixing with intermediary are compatible with existing cryptocurrency systems. But one common problem is that they require the same denomination, which means the group or the intermediary has to agree upon a predetermined transaction amount for everyone. Since the transaction value will be recorded on the blockchain, the payer-payee connection would be obvious, even under any mixing technique, if everyone sends a different amount of coins.

Therefore, alternative cryptocurrencies emerge to provide better anonymity and flexibility. For example, CryptoNote[20] is one of the first attempts to eliminate user coordination and provide large-scale privacy more efficiently. This idea, combined with confidential transactions and stealth addresses, is later developed to an alternative coin, Monero[23]. CryptoNote blinds transaction entries with one-time traceable ring signatures. It requires no cooperation among users and no trusted third-party. The transaction destination is a one-time public key that is derived from payee's published address and some randomness generated by the payer. The payer signs the transaction with the secret key, hides the corresponding public key for verifying her ownership among a chosen set of public keys, and then publishes the transaction information to the blockchain. With the nature of ring signature, the fund can come from any entity among the set of public keys. However, since CryptoNote heavily relies on ring signature, its anonymity set and performance are also limited by the size problem caused by ring signature. What's more, the scheme does not allow pruning on the blockchain, which means the blockchain has to remember everything from the beginning to the end.

     Monero is based on CryptoNote and obscures a transaction by adding mixins, previously spent coins of same denomination by other users, to the real input. So the external observers are supposed to be denied the ability to distinguish the real input from all coins included within a transaction. However, it is shown that the real coin spent is highly deducible and traceable because the anonymity in Monero is largely dependent on other users' previous transaction decision and the sampling of mixins[19]. First, 0-mixins, coins that are spent like a normal UTXO transaction in Bitcoin and are explicit about the transaction information, can also be included in a transaction when a user is trying to hide among coins. Second, when sampling mixins, the distribution is scattered among all past coins and real input is usually the newest transaction. Therefore, these coins can be instantly excluded from being the real input and the anonymity set becomes smaller than expected.

     ZeroCoin[15] is another early idea that later became the foundation of an alternative cryptocurrency. In a ZeroCoin protocol, the payer exchanges bitcoins for minted zerocoins. A zerocoin acts as a zero-knowledge proof of payments and is collected by a public cryptographic

accumulator. The owner of the zerocoin can later prove ownership for the transaction and claim the bitcoins to some address. Each zerocoin is attached to a serial number to prevent double spending. The link between the payer and payee is hidden by the zero-knowledge proof. There is no way to link a payer to a zerocoin since every zerocoin is minted at the transaction and there is no way to link a zerocoin redeemed to a payee since no information about the origin is contained in a zerocoin. ZeroCoin has limited efficiency due to its complexity and proof size. What's more, some operations are not supported by the Bitcoin script and the operations exceed the Bitcoin block size limit. Thus, ZeroCash[7] improves upon ZeroCoin and becomes a popular alternative cryptocurrency.

      ZeroCash is different from ZeroCoin in three ways. First, ZeroCash has two versions of addresses, one is transparent and the other is shielded. The shielded addresses in ZeroCash hide both payer and payee's addresses and the amount transferred. The transparent addresses and shielded addresses are allowed to interact. Second, it allows for different denominations and gives "changes" in a transaction. Each transaction mints coins and pours the transaction to the pool by proving the ownership of the new coins. Third, ZeroCash reduces the proof size of ZeroCoin and replaces the accumulator with a Merkle Tree. ZeroCash is no pruning, like Monero.

      Alternative coins are more flexible and allow more sophisticated cryptographic primitives for better anonymity. Another framework called Hawk[1] aims to build privacy-preserving smart contracts. It is not limited to providing fungibility in transaction, but works with any smart contracts. Hawk relies on the blockchain for accuracy and availability and trusts a "manager" for posterior privacy. There are many possibilities for new cryptocurrencies and complicated protocols. A major drawback is that the complexity of the protocols usually stops the majority of the community to trust and use alternative coins. Nevertheless, they usually work better with a larger crowd and derive stronger privacy guarantees from a trusted setup with a small group of people.

There are pros and cons in each category. Peer-to-peer on-chain mixing ensures k-anonymity for a transaction within a single mixing group, but it requires multiple rounds of communication and coordination among senders. Mixing with intermediary partially trusts a third party to facilitate the transaction, but the third party can become a single point of failure by colluding in some way or DoS the protocol. Alternative coins are more flexible and guarantee stronger anonymity, but they are significantly more complex and expensive. To balance the trade-offs between privacy guarantees, efficiency, and fairness, Eth-Tumbler is implemented as an alternative solution to a mixing problem.

**III. Eth-Tumbler**

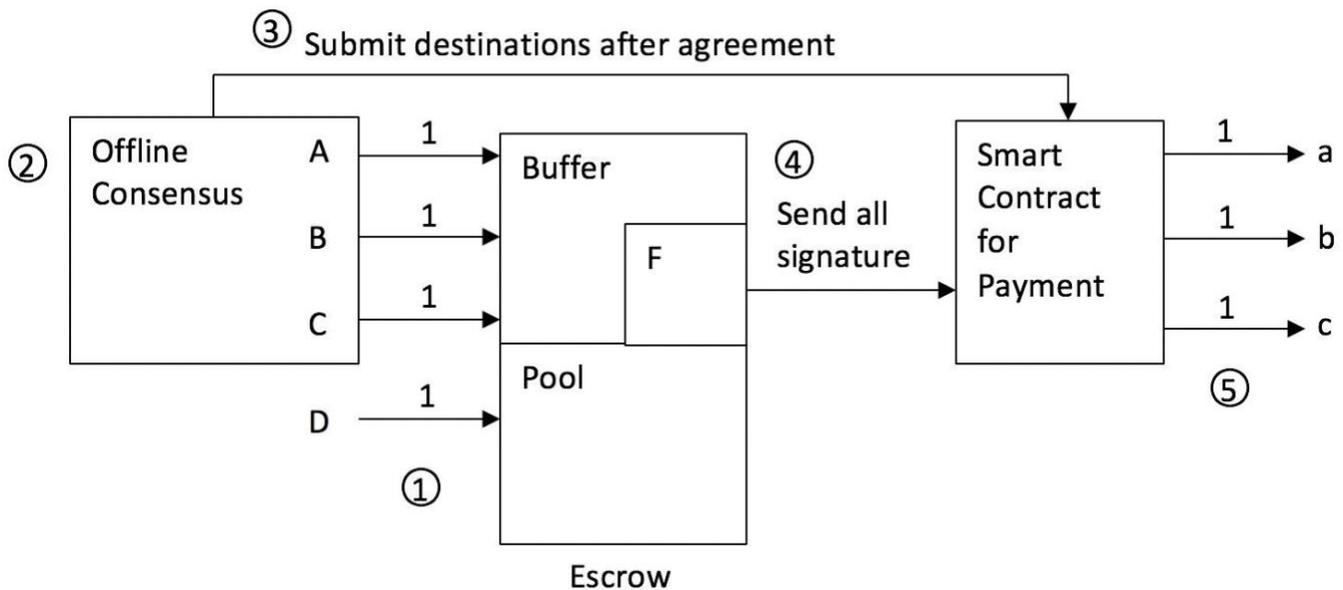

Figure 1. Eth-Tumbler mechanism

Eth-Tumbler is an implementation project that facilitates a mixing service and preserves user privacy. As shown in figure 1, each round of mixing consists of mainly five steps to complete.

**Step 1.** Each user sends a coin to the escrow and enters the buffer. When the escrow reaches its predetermined group limit, users who send money will wait in the pool until the transactions in the buffer have been resolved.

**Step 2.** When the buffer is full, users are connected by a chatroom channel and they perform layered encryption on their destination addresses. Each user first encrypts one address under every public key. After all encrypted addresses are posted to the channel, each user decrypts one layer for all encrypted addresses. The room ends with all destinations in the clear.

**Step 3.** The leader submits all destinations to the contract on the blockchain. Both the contract and the destinations should be checked for correctness.

**Step 4.** Every sender signs the message containing all destinations. The contract receives all destinations under multiple signatures. It first checks whether the number of destinations matches the number of signatures and the correctness of all signatures, then validates the payment for each signer.

**Step 5.** Once the escrow receives all signatures, it operates and forwards the escrowed amount in the buffer to the destinations submitted. One mixing round is thus complete, and the escrow "flushes" the current buffer and pushes users in the pool to the buffer for a new round.

Eth-Tumbler achieves the following properties that are ideal for a cryptocurrency exchange protocol:

- **Accountability.** This system prevents double-spending and ensures that the money eventually reaches a right destination. All funds directly go to the contract at first and are publicly verifiable on the blockchain. After that, the coins are either sent to payees or are retracted by the payers. The money won't be locked in the contract and will always be validated with the signature of its owner before being moved out from the contract.

- **Anonymity.** The payer is the only entity that knows his/her own payer-payee mapping. External observers cannot find out the links since the blockchain records the transactions as from the payers to the contract and the contract to the payees. Payees have the same view as external observers since they do not have to participate in the protocol to receive money. Payers are not aware of each other's destinations because of the layered encryption step. As long as the same denomination is required, k-anonymity within one round of buffer is guaranteed. The anonymity set cannot be reduced by traceability analysis, like the deanonymization attack on Monero in [19]. Because even though the anonymity is dependent on other users, there is no linkage to the information about users' previous spending behaviors by looking at this single transaction. In addition, the anonymity set makes the transaction all together so there is no timing difference on all inputs.

- **Compatibility.** This system as another peer-to-peer on-chain mixing protocol is compatible with Ethereum and many other programmable cryptocurrency.

- **Efficiency.** The anonymity set is scalable to the number of transactions that is possible to fit in a single block. In addition, this scheme requires no mixing fee.

- **Flexibility.** The users are provided with the flexibility to join and leave the mixing protocol at any point without losing money. The contract allows for withdrawing funds and leaving the group at any point of the protocol. Since the escrowed amount is locked in the contract and can be withdrawn, no one loses anything for joining or leaving except for a small amount of gas fee to run the contract.

- **Resilience.** Users are not affected by other people leaving the protocol. This system is not vulnerable to denial of service attacks under several scenarios as the source code is public. First, it is possible that no one takes action to go offline and do the mixing when the buffer is full. This would lead to the escrow stuck with the current status to wait for a signed destination message. However, anyone can start a new escrow and people who are stuck in any previous contracts can always withdraw the coins and join new contracts. Second, if someone deviates from the protocol in the middle of the mixing, malicious parties are forced out from the group so that the current round can continue with a smaller anonymity set. So even if the group consensus is not reached in one round, any user can choose to stay with the current group without suspicious collaborators or get out of that round and start a new one.

| Scheme | How Eth-Tumbler is better | How Eth-Tumbler is not better |
|---|---|---|
| CoinJoin | 1. No dependence on native multisig | 1. Simplicity<br>2. More possible variants for different functionalities |
| CoinSwap | 1. Unlinkable even to the mixer<br>2. Less transaction confirmations | 1. No cooperation among senders<br>2. Can happen across chain |
| MixCoin | 1. Mixer cannot steal or learn the links<br>2. Unlinkability among mixers<br>3. Resilience for DoS | 1. No cooperation among senders<br>2. Can happen across chain |
| TumbleBit | 1. No action required for payees<br>2. Withdrawable pre-deposit | 1. No cooperation among senders<br>2. Larger anonymity set |
| ZeroCash | 1. Cheaper and faster(zk)<br>2. Compatible with existing coins<br>3. Flexibility to enter and quit<br>4. pruning | 1. Different denomination(mint coin)<br>2. Less communication<br>3. No dependence on k-anonymity |
| CoinShuffle | 1. Solve key distribution<br>2. Flexibility to join and leave the group | 1. Came up with layered encryption earlier |
| Hawk | 1. No manager trusted for posterior privacy<br>2. Cheaper and faster(zk) | 1. Privacy-preserving smart contract (not limited to transactions)<br>2. Different denomination(hides amount through encryption) |
| CryptoNote | 1. No sender-receiver communication<br>2. Cheaper and faster(ring signature)<br>3. Pruning<br>4. No need for extra protection against double spending | 1. Allow a single public address<br>2. No cooperation among senders<br>3. Larger and more flexible anonymity set |

Figure 2. Eth-Tumbler Compared with some existing works

The biggest problems of Eth-Tumbler are also natural flaws for most peer-to-peer mixing schemes. First, it requires the same denomination within the same group, which means the amount of money to transfer has to be fixed. Second, it requires rounds of offline communication among payees. But overall Eth-Tumbler is simple to understand and flexible for users to join and leave. Therefore, the idea of Eth-Tumbler is comparable to the existing schemes in many ways.

## IV. Problems and Solutions

Below are more detailed descriptions about Eth-Tumbler with a focus on some problems I encountered during the implementation and our solutions.

**Public key distribution and group order**

For simplicity and efficiency, our implementation assumes that every user broadcasts a public key for encryption when entering the escrow. Alternatively, we can retrieve the wallet public key from the smart contract since users have made transactions for entering the escrow. The wallet address is not equivalent to the account public key in Bitcoin or Ethereum and is instead the hash. But the public key can be obtained when the account makes transactions[22]. Once we have the public key for each user, they can proceed with the layered encryption and signatures. In addition, we look at the last three digits of the public key of each user to determine the group order in the encryption chain. After they reach consensus, each user can compute new keys for signing future messages by themselves.

$$r \leftarrow \text{hash}(\text{Channel ID} + \text{pk\_enc})$$

$$\text{pk\_sig} \leftarrow g_{x+r} = g_x \times g_r \text{ where } g_x \text{ is the pk\_enc broadcasted}$$

**Blame**

In order to guarantee correctness and efficiency for the payment, we need a way to determine the suspicious users and force them out of the transaction group. Users can disturb the protocol in many ways. On one hand, they can participate in the buffer by sending a deposit to the escrow, but do not proceed with the communication. So the users that send in coins later and are waiting in the pool will be stuck in the escrow without being taken to the offchain channel. This can be prevented by adding a withdraw function for users to leave the current mixer and making the escrow code open-source and verifiable so that everyone can start a new valid mixer at any point.

*Scenario 1:*

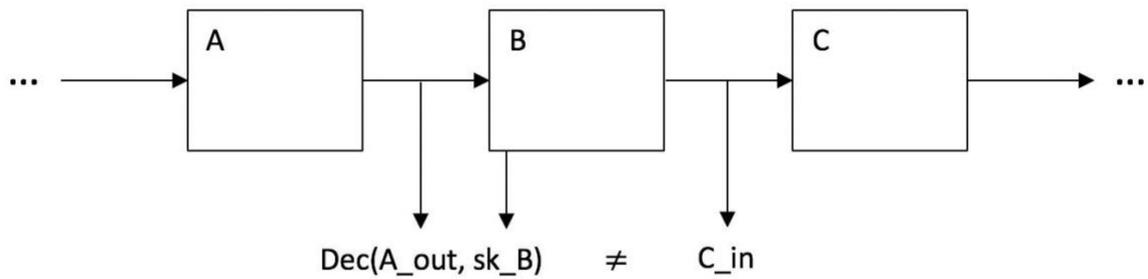

*Scenario 2:*

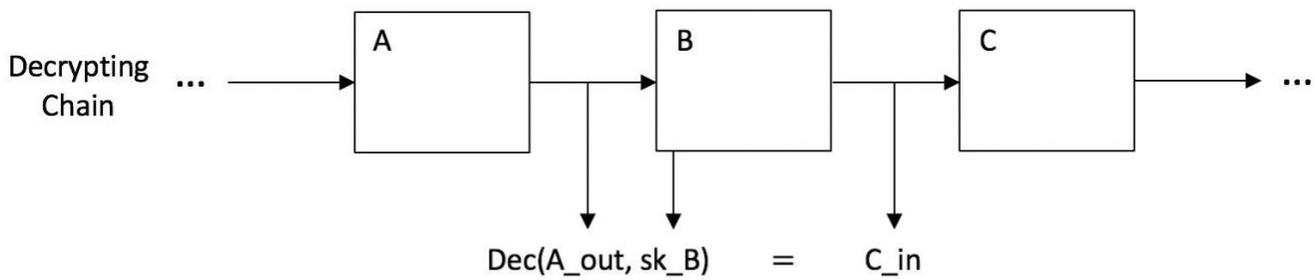

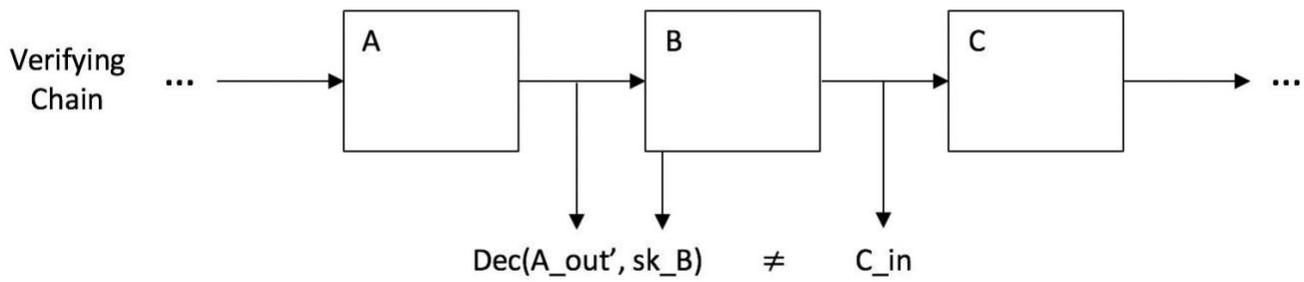

Figure 3. Eth-Tumbler Blaming Chain

On the other hand, users might drop or modify the destinations from the previous messages posted to the public channel while peeling off the layered encryption. To protect honest parties from losing money to wrong addresses, everyone can first check the destination addresses in clear after the round of layered encryption and decryption. If the number of users do not match the number of destinations or someone's address is not contained in the list, we then reveal all secret keys for all users. After that, it would be possible to find the parties to blame by decrypting each message on the layered decryption chain with each corresponding secret key, as shown in figure 3 scenario 1. Each user submits the in state, out state and secret key. If what should be posted does not match what has been posted at a certain position, the rest of the users will abandon the deviating parties at those positions and start a new round of mixing.

It is also possible that two users collude to falsely accuse a party between them of deviating, like scenario 2 in figure 3. In that case, either all three of them will be kicked out, or users in the middle proves in a certain time window that the other users are submitting different messages for the decrypting phase and blaming phase. This blaming method works since no one in the protocol will lose coins and people have the freedom to choose to continue with a reduced anonymity set or leave the group for a new one.

**Off-chain consensus with multiple signatures**

Since it is expensive to prove one's work and its correctness to the blockchain, we decide to reach consensus off chain and then sign the agreement on the blockchain to minimize the cost of the protocol. There are many alternative ways to sign a single message by multiple users, such as using an onion data structure or threshold signatures. However, for reducing the gas cost of the signature verifying functions in the smart contract, we choose to submit a concatenation of the signatures. After the layered encryption and decryption are finished and the destinations are checked, everyone signs the final destination by concatenating their generated tags under their signing keys. After the number of signatures match the number of destinations, the last user on the chain submits the message, all destinations with all signatures concatenated, to the smart contract. Our code uses Infura[16] endpoint for connecting to ethereum network and libraries "web3.js"[26] and "ethereumjs-tx"[10] to interact with smart contracts through pure Javascript. The smart contract will then verify the message against each tag under each corresponding public key. If all signatures validate, every group member's money in the escrow will be forwarded to their destinations.

## V. Conclusion

Eth-Tumbler is a mixing strategy that achieves great balance among security, privacy, efficiency, and fairness. It combines the merits of three main categories in the following way: it ensures on-chain k-anonymity within a transaction group, uses an escrow to pre-deposit, and allows for the flexibility to join and leave the protocol at any point without relying too much on other participants. Cryptocurrency fungibility is an interesting topic because it is difficult to behave for all users without the central registry and this decentralization adds extra dependence on other participants to perform any functionality. There is also a disagreement on whether these anonymity-enhancing protocols should be auditable to authorities or a custodian. Although this auditability seems to conflict with the assumption of strong user unlinkability, it is not unachievable and might become one possible direction for future research projects.